\documentclass[aps,prl,showpacs,twocolumn,english,superscriptaddress,floatfix]{revtex4}
\usepackage{graphicx}
\usepackage{dcolumn}
\usepackage{bm}
\usepackage{amssymb}
\usepackage{amsmath}
\usepackage{amsfonts}
\usepackage{amssymb}
\usepackage{color}

\begin{document}

\title{Experimental demonstration of quantum lithography beyond diffraction limit via Rabi oscillations}
\author{Jun Rui$^{1,2}$*, Yan Jiang$^{1,2}$*, Guo-Peng Lu$^{1,2}$, Bo Zhao$^{1,2}$, Xiao-Hui Bao$^{1,2}$, Jian-Wei Pan$^{1,2}$}
\affiliation{$^1$Hefei National Laboratory for Physical Sciences at Microscale and Department of Modern Physics, University of Science and Technology of China, Hefei, Anhui 230026, China \\ $^2$CAS Center for Excellence and Synergetic Innovation Center in Quantum Information and Quantum Physics, University of Science and Technology of China, Hefei, Anhui 230026, China \\ *These authors contributed equally to this work.}
\date{\today}

\begin{abstract}
Diffraction of light sets the fundamental limit for optical lithography. Many quantum lithography schemes have so far been proposed to overcome this limit either by making use of highly entangled photons, multi-photon processes or multiple Lambda transitions, which are all experimentally high-demanding. Recently, Liao et al. proposed a novel quantum lithography scheme which merely employs Rabi oscillation to surpass the diffraction limit. Here we report a faithful experimental realization of this scheme. Resolution up to ninth of the Rayleigh diffraction limit has been observed. Possibility of creating an arbitrary pattern is also tested experimentally by demonstrating the peak narrowing process using several Rabi floppings together with state-selective optical depletion. Our work may have direct applications in atom pattern engineering for quantum information or quantum simulation applications, and will also possibly boost the adoption of quantum lithography into real-world applications in the near future.
\end{abstract}

\maketitle

Optical lithography is the most important method to fabricate microscale structures in modern semiconductor industry~\cite{Ito2000}. The resolution of optical lithography is fundamentally limited by optical diffraction~\cite{Ito2000}. Many quantum lithography schemes~\cite{AlAmri2012409} have been proposed so far to create patterns with both the feature size and feature separation overcoming the diffraction limit, by either making use of the quantum nature of light or the quantum nature of matter. By making use of highly path-entangled NOON states~\cite{Boto2000, Angelo2001}, one can create interference patterns with a resolution which is enhanced by N times over the diffraction limit. Nevertheless, high-order NOON states with high intensities are extremely difficult to prepare. It is also possible to create sub-wavelength features by making use of multi-photon nonlinear processes with classical light~\cite{Agarwal2001,Hemmer2006,Bentley2004,Peer2004}, however the nonlinearity is typically rather small and extremely high laser intensities are required. There is also proposal which makes use of dark states in a multiple Lambda system and requires only moderate laser intensity~\cite{Kiffner2008}, nevertheless this scheme suffers from the difficulty of finding the well-structured multi-Lambda energy system and the adoption of many laser beams.

Recently, Liao \textit{et~al.} have proposed a novel quantum lithography scheme~\cite{Liao2010}, which simply involves two atomic levels and can in principle reduce the feature size and separation beyond the diffraction limit arbitrarily. This scheme relies on the process of Rabi oscillation driven by a standing-wave light field. The standing-wave field can be created by two-beam interference, thus giving an intensity period of $\lambda_{\mathrm{eff}}=\lambda/2\sin{\theta}$, where $\theta$ is half of the separation angle between the beams. As the Rabi frequency depends on the electric field strength, $\Omega_R=\vec{d}\cdot\vec{E}/\hbar$, nonlinearity of the light field is directly transferred to the spatial distribution of $\Omega_R$. At positions with maximal electric field strength, atoms oscillate fastest between internal states; while at positions of interfering minima, which ideally is zero, atoms can not be excited. If a $\pi$ pulse (for the peak $\Omega_R$) is applied, an atomic pattern with periodic spacing of $\lambda_{\mathrm{eff}}$ can be created. If a $N\pi$ pulse is applied, the atomic pattern resolution can be reduced to $\lambda_{\mathrm{eff}}/N$, which in principle has no upper limit. As a contrast, in traditional optical imaging and optical lithography, the Rayleigh criterion puts an upper limit of about $\lambda/2N.A.$ in resolution, where $N.A.$ is the numerical aperture. For lithography in air, this limit is $\lambda/2\sin{\theta}$, where $\theta$ here is half of the included angle of the objective lens. Thus, with the Rabi oscillation method, by applying a $N\pi$ pulse, resolution can be enhanced by $N$ times over the diffraction limit in classical optical lithography.

In this paper, we report the first experimental realization of this scheme. Two-photon Raman transitions are employed to induce Rabi oscillations between two internal ground levels. One Raman beam is periodically patterned through interference, which gives $\lambda_{\mathrm{eff}} \simeq 57\mu$m, while the other is flat for simplicity. We manipulate laser-cooled atoms with these Raman beams, and make use of a specially designed absorption imaging system to measure the atomic patterns. High-resolution spatial patterns up to ninth of the Rayleigh diffraction limit have been created and observed for atoms in specific hyperfine state. Moreover, by combining the Rabi oscillation together with state-selective optical depletion, arbitrary patterns can be produced in principle \cite{Liao2010}. In our experiment, we demonstrate the process of peak narrowing without changing the peak separation by applying multiple pulses.

\begin{figure}[htb]
\includegraphics[width=\columnwidth]{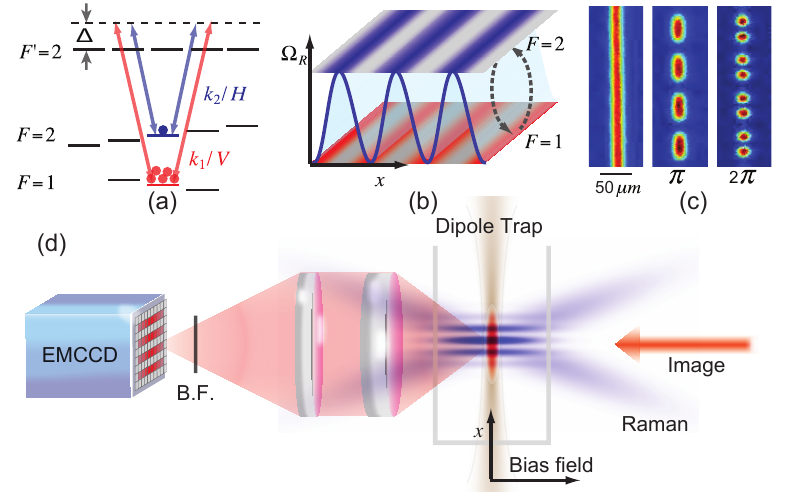}
\caption{Experimental scheme. (a) Energy levels. The $D1$ line transition of $^{87}$Rb is used as the channel for two-photon stimulated Raman transitions, as well as for $\pi$ pumping. But for absorption imaging, $F=2\to F'=3$ cycling transition of $D2$ line is used. (b) Illustration of spatial distribution of Rabi frequency and atomic pattern on each hyperfine ground state. (c) Typical atomic patterns in experiment. Left: in the dipole trap; Middle: after a $1\pi$ Raman pulse. Right: after a $2\pi$ pulse. (d) Experimental layout. A cold atomic ensemble is captured with a dipole trap inside the vacuum glass cell. Then a simple interferometer with two arms is placed outside the vacuum, where Raman beams are splited and polarization rotated, then interfere at the center. After the cell, a high resolution lens group compensates the abberation introduced by the glass cell and then a $780$~nm narrow bandpass filter (B.F.) is placed before the CCD camera to filter out other scattered photons except the imaging beam.
}
\label{fig1}
\end{figure}

The detailed experimental scheme is shown in Fig.~\ref{fig1}. A cold $^{87}$Rb atomic ensemble is prepared by first capturing with a standard magneto-optical trap (MOT) and then a molasses cooling process which results in an atom temperature of $\sim10~\mu$K. Afterwards, the atoms are loaded into an optical dipole trap ~\cite{Kuppens2000} to compress the radial dimension of the atomic ensemble.  The dipole trap is provided by a tightly focused $850$~nm laser beam from a diode laser, with a waist of about 30~$\mu$m and $170$~mW power, thus giving a trapping depth of about $57~\mu$K. The atoms held in the dipole trap have a width of about $27~\mu$m and peak optical depth of about $0.3$, as shown in Fig.~\ref{fig1}(c). Before driving Rabi oscillations, these atoms are released from the dipole trap, and then optically pumped to the state of $|F=1, m_{F}=0\rangle$. To image the projected atom patterns down to several $\mu$m, we use a high resolution imaging design similar to Ref.\cite{Mun2008}, to compensate the spherical aberrations from the vacuum glass cell. The imaging system has a magnification of $9.7$ and a resolution better than $4.5~\mu$m, as measured with a Thorlabs test target. An Andor IXon EMCCD, with pixel size of $16~\mu$m, is used to record images for off-line analysis. Absorption imaging is taken with the $D2$ line cycling transition of $^{87}$Rb atoms, where only atoms on $F=2$ ground state are detected.

To create the standing-wave light fields for driving Rabi oscillations, we modify the original two-level scheme to a three-level stimulated Raman transition process~\cite{moler92}. As shown in Fig.~\ref{fig1}(a), two hyperfine ground sublevels are connected via a coherent two-photon transition process, where the Raman fields are provided by two phase locked diode lasers, and each with a single photon detuning much larger than the nature linewidth of $D1$ line, $\Delta\gg\Gamma$. For $^{87}$Rb atoms, there are two possible paths for two-photon transitions between $|F=1, m_{F}=0\rangle$ and $|F=2, m_{F}=0\rangle$ ground sublevels. Given two Raman fields with H and V polarizations, constructive interference will happen between these two paths due to the Clebsch-Gordan coefficients of relative transitions, thus two-photon transitions can still be driven between these two sublevel states.

The layout of the Raman fields is shown in Fig.~\ref{fig1}(d), where one of the Raman beams is splitted into two paths by a Glan-Taylor prism, and then the polarization of the reflected path is rotated back to be horizontal to make them interfere at the center, leaving a standing-wave light field there. The other beam, however, only transmits through one path of the prism and forms an homogeneous field at the center. The two-photon Raman Rabi frequency depends on single-beam Rabi frequencies, $\Omega_{R} (x) \propto \Omega_{1}(x)\Omega_{2}(x)\propto E_{1}(x) E_{2}(x)$, where $\Omega_{i}(x)= \vec{d}\cdot\vec{E}_{i}(x)/\hbar$ and $i=1,2$ correspond to light fields of Raman beam $1$ and $2$. Note the ideal interfering field is $E_{1}(x)=2E_{1}^{0}\cos{\pi x/\lambda_{\mathrm{eff}}}$ and the flat field is $E_{2}(x)=E_{2}^{0}$, thus Raman Rabi frequency can be described as $\Omega_{R}(x)= \Omega_{R}^{0}\cos{\pi x/\lambda_{\mathrm{eff}}}$, where $\lambda_{\mathrm{eff}} \simeq 57~\mu$m for half separation angle of $\theta=0.4^{\circ}$. Note if both beams are splitted into two paths and then interfere with themselves at the center, relative path length control of the interfering arms is necessary to make sure both interfering patterns are in-phase or with a special phase offset. And due to the simplified configuration here, atoms on $F=1$ state experience position dependent light shifts, $\Delta^{1}_{AC}(x)\propto\Omega^{2}_{1}(x)/\Delta$, while atoms on $F=2$ have homogeneous shifts. This spatial periodic differential light shifts, created along the axial direction of atoms in the dipole trap, will limit the spatial visibility at some positions but will not affect the resolution enhancement according to our simulations.

\begin{figure}[htb]
\includegraphics[width=\columnwidth]{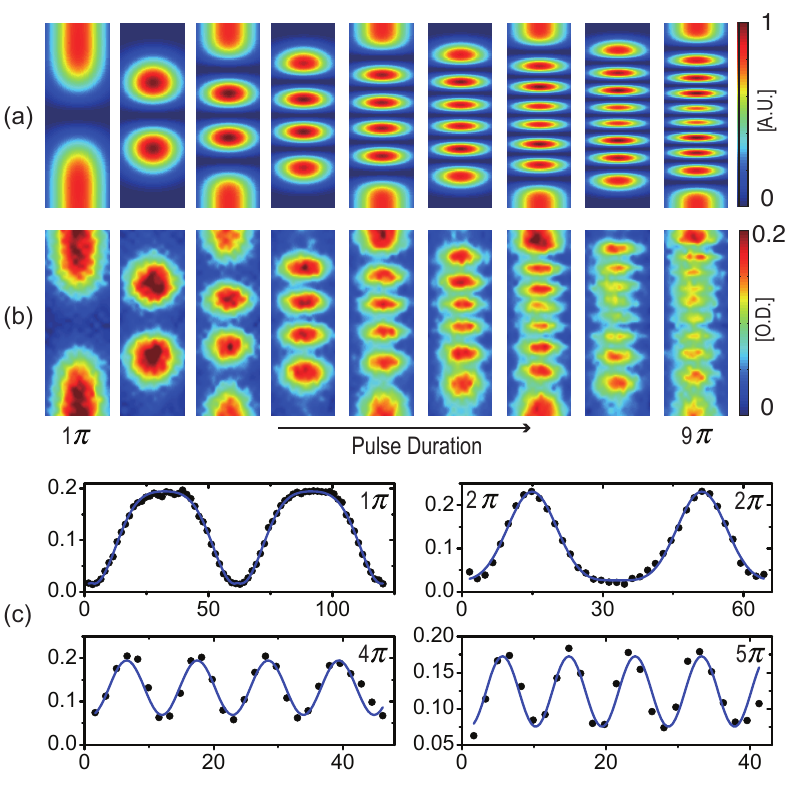}
\caption{Resolution enhancement with various Raman pulse durations in a single $\lambda_{\mathrm{eff}} $ range. (a),(b) Simulated and experimental atomic patterns in a single $\lambda_{\mathrm{eff}}$ range. The measured pictures are interpolated with 4 points in each pixel area to show the patterns clearly. It's demonstrated with a pulse duration of $9\pi$, up to $9$ peaks can be created in the single intensity period area.
(c) Fitted pattern visibility for several pulse durations, where the horizontal axis is position in space with unit of $\mu$m, and vertical axis is optical depth. These points used in fitting visibility are raw pixel points. Visibilities for $1\pi$, $2\pi$, $4\pi$ and $5\pi$ durations, are $0.83$, $0.79$, $0.48$, and $0.39$ as fitted, respectively.}
\label{fig2}
\end{figure}

In our experiment, the Raman beams have a diameter of $1.3$~mm, each has a single photon detuning of $\Delta=+750$~MHz relative to $F'=2$ state of the $D1$ line. Raman beam 1 has a total power of  $1$~mW, while beam 2 has a power of $2$~mW to match the light shifts for atoms at interference peak positions. With these beams, the on resonance Raman Rabi frequency at peak positions is calculated to be about $2\pi\times550$~kHz. The two-photon detuning frequency is set to be $240$~kHz, which is to match the on resonance condition for atoms at half of the maximum Rabi frequency positions. Taking account of detuning, the spatial maximum Rabi frequency becomes $2\pi\times 605$~kHz, which agrees well with our measurements. In Fig.~\ref{fig1}(c), periodic atomic patterns clearly appear after applying $1\pi$ or $2\pi$ pulses with duration of $0.8~\mu$s or $1.6~\mu$s, respectively. Considering the pixel size and magnification ratio, the periodic separation is fitted to be $\lambda'_{\mathrm{eff}}=59.3~\mu$m for $1\pi$ pulse, which corresponds to the diffraction limit if projecting an image through this small aperture.

After the initial exploration with short $\pi$ pulses in Fig.~\ref{fig1}(c), we now turn to multi-$\pi$ pulses. To present the structures even clearer, we now focus on the area of a single effective wavelength range, as shown in Fig.~\ref{fig2}(a-b). It clearly demonstrates that up to nine peaks can be created in a single $\lambda_{\mathrm{eff}}$ range, where the smallest separation reaches down to about $6.5~\mu$m. For short pulse durations up to $5\pi$, we find it's able to fit the pattern distribution with normal sinusoidal functions very well as shown in Fig.~\ref{fig2}(c), and the visibility is calculated as the amplitude of oscillation divided by the average. For longer pulse durations, although it's able to identify number of peaks, the visibility gets reduced severely. This is mainly contributed by the relative phase fluctuations between two interfering arms. Second, atoms are heated up during releasing from the trap and optical pumping. Diffusion of atoms will finally erase the atomic patterns for several $10~\mu$s, thus the patterns have to be imaged immediately after creation.

\begin{figure}[htb]
\includegraphics[width=\columnwidth]{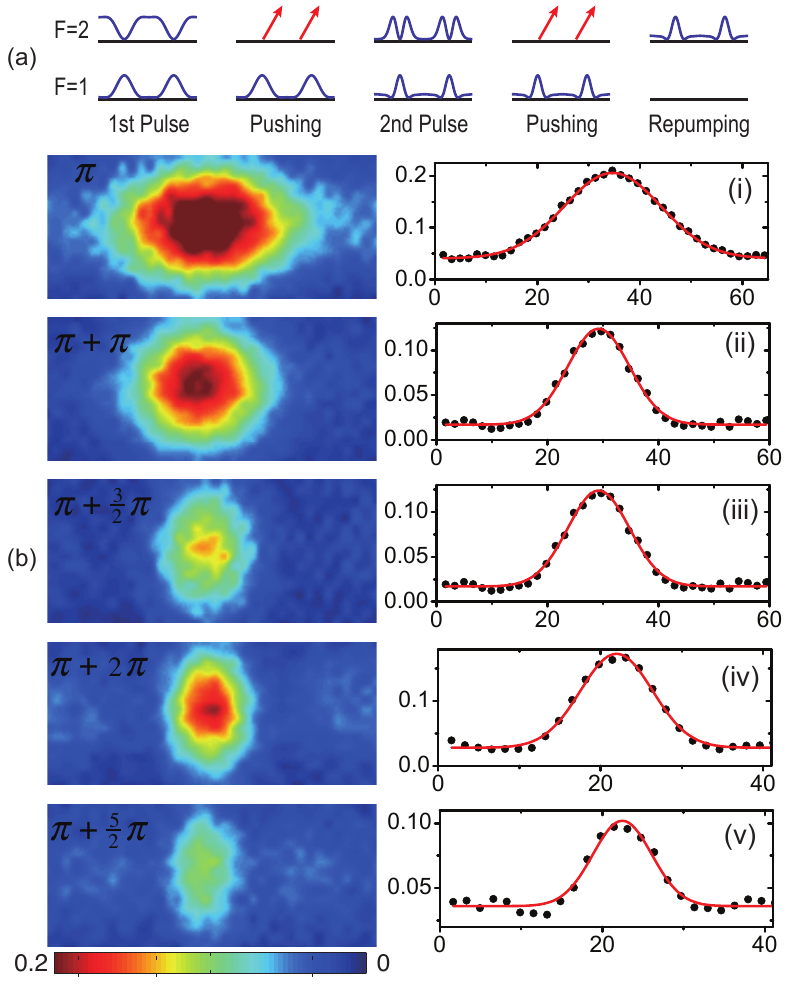}
\caption{Narrowing of the atomic pattern with different pulse combinations. (a) Sequence of the experiment. Atomic patterns are first created on $F=1$ state, and then narrowed with the second Raman pulse. Then after all atoms originally on $F=2$ state are pushed away of the area, the patterns are repumped to $F=2$ for imaging. (b) Experimental results of width narrowing, with the first pattern being created by a $\pi$ pulse. Left, absorption imaging of the cloud, images here are rotated relative to in Fig. \ref{fig2}; Right, cross-section view of the axial width, where horizontal axis denotes the position in real space with unit of $\mu$m and vertical axis denotes optical depth. For specific pulse combinations and peak widths, see in text.}
\label{fig3}
\end{figure}

With only a single Raman pulse, we have already shown the diffraction limit can be easily circumvented via Rabi oscillations driven by a standing-wave field. With a combination of several pulses, there are more exciting prospects with this method, i.e., creating arbitrary patterns as proposed in the original scheme \cite{Liao2010}. Here we show how to create special patterns with two pulses, i.e. narrowing the peak width but without altering the separation distance, as shown in Fig.~\ref{fig3}. All atoms are initially prepared on $|F=1,m_{F}=0\rangle$ state, then a $\pi$ pulse is applied to create the initial atomic pattern on $F=1$ state. Subsequently, a strong depleting beam resonant with the $D2$ line cycling transition is applied in vertical direction during $20~\mu$s, thus pushing all atoms on $F=2$ state out of the imaging area. Afterwards, the second Raman pulse with specific duration is applied, to narrow the peak width of atoms on $F=1$ state. Then, the pushing beam with $20~\mu$s duration is shined again to clear all atoms transferred to $F=2$ state. In the end, a repumping beam is applied to transfer the spatial distribution from $F=1$ to $F=2$ state for absorption imaging. The experimental results are shown in Fig.~\ref{fig3}(b), where the second Rabi pulse with durations of zero, $1\pi$, $\frac{3}{2}\pi$, $2\pi$, $\frac{5}{2}\pi$ are used. For each duration, $20$ pictures are averaged for analysis. The fitted widths for these pulses are found to be $38.4~\mu$m, $37.8~\mu$m, $31.2~\mu$m, $22.3~\mu$m, $17.8~\mu$m, and $14.5~\mu$m, where the smallest width realized here is about  $1/4$ of the diffraction limit. Note that, before taking these images here, atoms on $F=1$ state have already thermally diffused, during pushing away and repumping processes. As a reference, the theoretical predicted widths for these pulse combinations are, $31.0~\mu$m, $24.6~\mu$m, $18.5~\mu$m, $14.0~\mu$m, and $12.8~\mu$m, respectively, where the slightly larger widths measured can be explained as above.

There are several technical imperfections yet to be resolved in future works. First, with traditional optical imaging, the detection itself is diffraction limited, that's why we have to explore the diffraction nature at scales much larger than optical wavelength. To fully explore the ultimate performance of the scheme for grazing incident Raman beams, at the scale of subwavelength,  other methods like atomic force microscope have to be utilized to measure after imprinting the pattern on a substrate. Second, relative phase of the interference arms hasn't been stabilized in this work. Without phase control, the interference field randomly drifts at the center at the scale of minutes, which is the same time scale of taking pictures for averaging in the experiment. This limits the maximal resolution can be achieved in our current work. Third, to simplify analysis and control, only single Raman beam is splitted and then interfering at the center. This results in inhomogeneous differential Stark shifts between internal levels across the standing-wave area, and limits the uniformity of visibility across the sample. The inhomogeneous shifts can be cancelled by making both of the Raman beams interfere at the center as well as making sure both standing waves are in-phase with each other. Finally, to create arbitrary patterns as proposed in the original scheme, multiple Raman pulses with different durations, controllable node positions of the standing wave, as well as a continuous incoming atomic source are necessary.

To summarize, the experiment presented here provides a clear demonstration of the quantum lithography scheme based on Rabi oscillations, and shows that resolution can be enhanced by up to nine times over the diffraction limit. With multiple Raman pulses, we can also create much narrower atomic peaks without altering the peak separations, where the smallest width is about one-fourth of the diffraction limit. Although patterns created are only one dimensional here, it's straightforward to extend this scheme to two-dimension, only by using the other Raman beam to create a standing-wave field along the other direction, to create a two-dimensional Rabi frequency field. In comparison with the traditional atom lithography experiments \cite{Timp1992, Johnson1998, Muetzel2002} or recent atom localization experiments \cite{Li2008, Proite2011, Miles2013} with dark states, our experiment highlights the advantage that both the feature size and feature separation can overcome the diffraction limit. Our work may have applications in optical lithography, atom pattern engineering for quantum information or quantum simulation experiments, and could also possibly boost the adoption of quantum lithography into real-world applications in the near future.

We thank M. Suhail Zubairy for stimulating discussions. This work was supported by the National Natural Science Foundation of China, National Fundamental Research Program of China (under Grant No. 2011CB921300), and the Chinese Academy of Sciences. X.-H. B. and B. Z. acknowledge support from the Youth Qianren Program.

\bibliography{Highrefs}

\end{document}